\documentclass[letterpaper,reqno]{amsart}
\usepackage{setspace}
\usepackage{amsmath,latexsym,amsbsy,amssymb,color,subcaption}
    \usepackage{ifpdf}
    \ifpdf
      \usepackage[pdftex]{graphicx} 
      \usepackage[pdftex]{hyperref}
    \else
      \usepackage[dvips]{graphicx}  %
      \newcommand{\href}[2]{#2}
    \fi

\newtheorem{definition}{Definition}[section]  

\newtheorem{R}{Remark}[section]

\newcommand{\h}{\hspace*{.24in}}
\def\T{{\mathcal T}}
\def\var{\operatorname{Var}}
\def\V{\operatorname{Vol}}
\newtheorem{theorem}{Theorem}[section] \newtheorem{lemma}{Lemma}[section]

\def\RR{\mbox{$I\hspace{-.06in}R$}}

\begin{document}
\makeatletter	   
\renewcommand{\ps@plain}{%
     \renewcommand{\@oddhead}{\textrm{Your Header}\hfil\textrm{\thepage}}%
     \renewcommand{\@evenhead}{\@oddhead}%
     \renewcommand{\@oddfoot}{}
     \renewcommand{\@evenfoot}{\@oddfoot}}
\def\specialsection{\@startsection{section}{1}%
  \z@{\linespacing\@plus\linespacing}{.5\linespacing}%
  {\normalfont}}
\def\section{\@startsection{section}{1}%
  \z@{.7\linespacing\@plus\linespacing}{.5\linespacing}%
  {\normalfont\scshape}}
\makeatother

\title[Experimental Design for Dynamics Identification]{Experimental Design for Dynamics Identification of Cellular Processes}         
\author{Vu Dinh, Ann E. Rundell and Gregery T. Buzzard}
\thanks{Vu Dinh and Gregery T. Buzzard are with the Department of Mathematics, Purdue University}
\thanks{Ann E. Rundell is with the Weldon School of Biomedical Engineering, Purdue University}
\thanks{All authors partially supported by NSF grant DMS-0900277.}

\begin{abstract}
We address the problem of using nonlinear models to design experiments to characterize the dynamics of cellular processes by using the approach of the Maximally Informative Next Experiment (MINE), which was introduced in  [W. Dong, et al. Systems biology of the clock in neurospora crassa. {\em PLoS ONE}, page e3105, 2008] and independently in [M. M. Donahue, et al. Experiment design through dynamical characterization of non-linear systems biology models utilising sparse grids. {\em IET System Biology}, 4:249--262, 2010].  In this approach, existing data is used to define a probability distribution on the parameters; the next measurement point is the one that yields the largest model output variance with this distribution.  Building upon this approach, we introduce the Expected Dynamics Estimator (EDE), which is the expected value using this distribution of the output as a function of time.  We prove the consistency of this estimator (uniform convergence to true dynamics) even when the chosen experiments cluster in a finite set of points.  We extend this proof of consistency to various practical assumptions on noisy data and moderate levels of model mismatch.  Through the derivation and proof, we develop a relaxed version of MINE that is more computationally tractable and robust than the original formulation. The results are illustrated with numerical examples on two nonlinear ordinary differential equation models of biomolecular and cellular processes.
\end{abstract}      

\maketitle


\textbf{AMS subjects classification.} 62K20, 65D15, 62L05, 92C45.

\section{Introduction}
The development and simulation of mathematical models of cellular processes can enhance our understanding of the underlying biological mechanisms (\cite{Kitano}).  Two important components of model development are the collection of data and the tuning of parameters for a given model structure to approximate the data.  In many settings, the collection of data is difficult and/or expensive, while tuning model parameters to data often involves a difficult nonlinear optimization, with potentially many local optima.  Moreover, the choice of data may make this tuning more or less difficult; thus we aim to design experiments to collect the most informative data for a given model structure.  

The review \cite{kreutz-timmer} provides a broad overview of model-based experimental design methodologies for systems biology, including methods for various optimality conditions governing unique, structural and practical parameter identification.  Of course, many books and articles have been written about experimental design, both from the frequentist and the Bayesian points of view.  We make no attempt to review them all here; a classic mathematical reference is \cite{pukelsheim}.  Many methods for experimental design focus on identifying the parameters -- designing experiments to minimize some measure of uncertainty in the parameter values given a model structure.  

In contrast, we are more concerned with developing a method to explore and elucidate the observable response (which we refer to as the output dynamics) of a cellular process rather than identifying the model parameters themselves. One motivation for this is that for a systems biology model with $N$ parameters, the set of possible output dynamics is often contained in a space of lower dimension $l \ll N$ (or perhaps in a small neighbourhood of such a space). This feature, which is an obstacle for the problem of unique parameter identification, is an advantage for designing experiments to identify dynamics: that is, we can choose a good design with very few experiments (approximately $O(l)$), but still obtain enough information to identify the dynamics.

Methods related to approximating the observable response as a function of independent input variables fall under the broad heading of regression or response surface methodology.  Once again there are many books and articles on the topic of experiment design for fitting response surfaces, and there are many approaches for representing a response surface. Most such approaches (e.g., Kriging and generalized polynomial chaos) seek to approximate the response surface with a linear combination of a fixed set of basis functions, such as polynomials or trigonometric functions.  See \cite{myers} for an overview.   

In this paper we focus on experiment design for accurate approximation of the response surface using a given biologically-based model. However, beyond this, as explained in \cite{D}, the method we apply acts as a kind of imaging method for understanding the behavior of a cell.  Based on an initial understanding of cell behavior (which is likened to an image from a microscope), we choose the next experiment to provide as much resolving power as possible in our next measurement 
(which is likened to focusing the microscope to enhance a particular feature). Instead of a linear combination of basis functions, as is often assumed in the experiment design literature, we assume the model structure encodes the dominant interactions and mechanisms using a nonlinear system of differential equations. To fix ideas, suppose our model output has the form $y = f(\omega, t)$, where $\omega \in \Omega \subset \RR^N$ is a fixed vector of unknown parameters, and where our quantity of interest is the dynamics output, which is a function of time, $t \in [0, T]$ (more generally $t$ could be a vector of inputs to represent any independent variables such as time, voltage, etc.). Measurements of $y$ at a given time $t$ can be modeled as a random variable. An estimate of $\omega$ based on these random variables is also a random variable. This estimate can be used to estimate the output $y(t)$ for any $t$, again giving a random variable. Classical experiment design typically seeks to minimize the variance in the estimate of $\omega$ or $y$. One approach to designing experiments for accurate response surface modeling is to use the condition of G-optimality. Roughly, this condition chooses an experiment design to minimize the variance in the output. In the case of a model that is linear in parameters, the Kiefer-Wolfowitz Equivalence Theorem states that this is equivalent to D-optimality, in which the design is chosen to minimize the determinant of the  covariance matrix for the estimate of $\omega$ (the inverse of the Fisher Information Matrix) \cite[Chapter 9]{pukelsheim}.  There is an extension of this result to nonlinear models \cite{white}; however, this result depends upon knowing the true parameters in the model, which are not known in general.   In \cite[Section 5.6]{federov}, this problem is addressed by either (i) using a minimax approach, in which the design is chosen to minimize over experiment designs the maximum over parameter space of the determinant; or (ii) using a Bayesian approach, in which an optimality criterion for a design (such as the determinant of the dispersion matrix or the maximum output variance) is averaged using a prior distribution on parameter space, and then the design is chosen to maximize this expected criterion.  A computational difficulty with both of these approaches is the need to evaluate a complex optimality criterion at many points in parameter space for each candidate design.  

Alternatively, the Maximally Informative Next Experiment (MINE) algorithm proposed in \cite{D} and later \cite{M} uses a sequential approach to experiment design in which existing data is used to construct a probability distribution on the parameter space; this distribution is then used to calculate the variance in the output as a function of time (perhaps normalized by expected experimental variance).  Based on this calculation, the next measurement will be taken at the time point with highest current (normalized) variance.  That is, the next sampled time point will be chosen at the time point that has highest current uncertainty in the output.  This method of design is modified to produce a parallel (nonsequential) design in \cite{J}.  Intuitively, each new point in such a design should provide the maximum possible information about the dynamics of the system and hence lead to convergence to the true system dynamics.  This approach is theoretically appealing in that it doesn't depend on an estimate of the true parameter values, and it is computationally appealing in that it requires a relatively simple sampling over the parameter space according to a specified distribution; this may be achieved reasonably efficiently with Monte Carlo methods. 

However, little is known about the convergence properties of this method:  Is this method sufficient to characterize the response surface (in the limit as the number of experimental points tends to $\infty$)?  In fact, in general, this scheme will not sample densely over the interval $[0,T]$, so it is not at all clear that it is sufficient to completely characterize the dynamics over this interval.  Moreover, it's not entirely clear how to use these nondense samples to estimate the dynamics.  

Motivated by the MINE algorithm, we address the following two problems for the identification of systems dynamics:
\begin{enumerate}
	\item[(A)] Specify the Dynamics Estimator: Given a set of data $(t_1,d_1), (t_2,d_2),...,\linebreak (t_m,d_m)$ and a model $y = f(\omega, t)$, how should we estimate the system dynamics? 
	\item[(B)] Prove convergence of Dynamics Estimator: For a given sequential approach to choosing measurement points $t_j$ and given the dynamics estimator in (A), do the estimated dynamics converge to the true dynamics?  
\end{enumerate}
In the derivation of the solutions to these problems, we developed variations of the MINE algorithm that are more computationally efficient than the original.  We describe these variations and solutions to Problems (A) and (B) with various assumptions in the body of the paper.  

Most approaches to problem (A) use the data to estimate parameter values and then use these parameters to obtain the corresponding dynamics.  For a complex, nonlinear model this is a difficult optimization problem with possibly many local optima and perhaps even multiple global optima.   In place of using an estimated vector of parameters to estimate the dynamics, we propose what we call the expected dynamics estimator (EDE).  This uses the available data to induce a probability distribution on parameter space and then averages the dynamic output using this distribution.  There are a number of advantages of this method of dynamics identification over parameter identification. First, since the dynamics for a deterministic system are unique, we don't need to worry about multiple global solutions. Second, by using the EDE, we look for the average behaviour of the system (with respect to a carefully constructed probability distribution). This task is typically much simpler than solving a nonlinear optimization problem. Furthermore, the Markov Chain Monte Carlo method can be employed to reconstruct the system's true dynamics. 

Another important advantage of the probabilistic framework over parameter estimation (via optimization) is that it is a feasible approach in cases of unidentifiability. A crucial problem in parameter estimation is the calculation of confidence intervals for the estimated parameters. In the simplest scenario when the number of data points is smaller than the number of parameters, any parameter estimation (via optimization) method will fail to provide a reliable estimate of the confidence region. Such methods (that return a single parameters estimate) will never be able to predict unknown output with high confidence (or any confidence at all). In order to do so, it needs to compute all possible parameter values that are consistent with available data, which is very unlikely in practice. This also extends to the case when the model's parameters are unidentifiable, which is a common phenomenon in systems biology. Our probabilistic framework provides a feasible way to address this issue: a given set of measurements gives a probability distribution on parameters, which can be used to construct confidence interval for output dynamics in addition to the EDE.

Problem (B) is a question about the consistency of the estimator (the ability to recover the true dynamics) as a function of a particular choice of measurement points. This question highlights the fact that the ability of the EDE to recover the true dynamics (consistency) depends heavily on the experimental design algorithm.  We note here that although the MINE method shares some features of a Bayesian approach, in that a probability distribution on parameters is updated based on new data samples, it does not fall into the class of Bayesian experimental design since the design points are not chosen to maximize an expected utility function.

\bigskip
This paper is organized to prove and illustrate the consistency of the EDE in various situations that progressively increase in complexity towards practical applicability. In Section~\ref{mathFramework}, we introduce the mathematical framework and the main assumptions about the behaviour of the investigated model that we use throughout the paper.  We also define the EDE to address problem (A). Section~\ref{EDEConsistency} addresses problem (B) for the ideal case when the investigated model is a correct model (can reproduce the true dynamics exactly) and data are noiseless. Theorem \ref{Theo1} deals with the case in which the experiments are made at random time points;  this result is provided primarily to illustrate the ideas to be used in later results but in a setting that avoids some technical assumptions that are needed later. Theorem~\ref{Theo2} and Theorem~\ref{ThmDiscrete} provide results in the case when the experiments are designed sequentially as in \cite{M} and \cite{D} in two different settings: when the parameter space is discretized and when the set of possible measured time points and output values are discretized. We then extend the consistency result to a larger class of designs by relaxing the choice of a point with maximal variance to a point with variance within a fixed constant multiple of maximal variance (Theorem~\ref{ThmWithC}). Our results imply that for these designs, we can always recover the true dynamics, even if all the measurements are made in a small portion of the time interval $[0,T]$. Section~\ref{noisyData} extends the result about EDE's consistency to the case when the experimental data are subject to random noise (Theorem \ref{Theo5}).  In this section we require that the set of possible measurement points is finite in order to guarantee convergence even in the face of noisy data.  This assumption is reasonable for the practical implementation of any experiment design.  Section~\ref{modelMismatch} relaxes the requirement of a correct model by allowing for a bounded error between the true dynamics and the closest approximation of the model (Theorem \ref{Theo6}).  From this result, we also justify the use of approximation methods in the algorithm to design experiments. In Section~\ref{sec:examples}, we illustrate our theoretical findings and demonstrate the efficacy of our method to design sequential experiments for dynamics identification with various biological models.  We also give an example to show that the choice of a design point within a fixed constant of maximal variance can lead to a faster rate of convergence of the EDE relative to the original MINE algorithm.  It is worth noting that although the framework we use in this paper is sequential, one can extend the result to the parallel case following the approach suggested in \cite{J}.

\section{Mathematical framework} \label{mathFramework}
\subsection{Model formulation}
We assume a mathematical model of a cellular process in the form
\begin{align*}
\dot{x}&=\alpha(\omega,x) \mbox{\hspace{.3in}} \text{(System of ODEs)}\\
x(0)&=x_0(\omega) \mbox{\hspace{.4in}} \text{(Initial conditions)}\\
y(t)=f(\omega,t)&=\beta(\omega,x(t)) \mbox{\hspace{.3in}} \text{(Output)}
\end{align*}
where $x=(x_1,x_2,...,x_{n_x})\in M \subset \RR^{n_x}$ is the state variable, with M a subset of $\RR^{n_x}$ containing the initial state, and $f(\omega,t) \in \RR^L$ is the observable response (output dynamics) that correspond to $L$ different experimentally observable quantities. Throughout this paper, for the sake of simplicity, we will assume that $L=1$. However, all of the arguments can be extended to the case of multi-dimensional output without any difficulty.

It is worth noting that the set of possible outputs is not necessarily the same as the number of dynamic variables occurring in the system. An output could be any kind of prediction, e.g. also a sum or ratio or even integral of dynamic variables. However, in the case $L=1$, there is only one observable output $y$. Identification of $y$ will lead to identification of all possible outputs as well as a characterizaton of the uncertainty in unidentifiable outputs.

The purpose of our experimental design framework is to determine as accurately as possible the output dynamics based on measurements. This is a kind of interpolation problem. We do not address the extrapolation problem, in which measurements of one output are used to make inference about an unobservable quantity.

The vector of unknown parameters is denoted by $\omega=(\omega_1,...,\omega_N) \in \RR^N$ and is assumed to belong to a subset $\Omega$ of $\RR^N$. In most parts of the paper, { the parameter space} $ \Omega$ will be assumed to be an open set along with a probability measure on $\Omega$, or a discrete subset of $\RR^n$ along with a probability measure. The components of $\alpha$ and $\beta$ are assumed to be $C^1$ functions of their arguments. These functions and initial conditions may depend on the parameter vector $\omega \in \Omega$.

%

The system will therefore be associated with the mapping $F:\Omega  \to C^1([0,T],\RR)$ defined by $F(\omega) = f(\omega,\cdot)$, where $f(\omega,\cdot)$ is the observable response of the system as a function of $t \in [0,T]$ for a given $\omega$.  
The image of $\Omega$ under $f$, $Y=f(\Omega, \cdot) \subset C^1([0,T],\RR)$ will be referred to as the dynamics space in this paper.

Throughout this paper, the true dynamics and the data values at a given time, $t$, will be denoted by $g(t)$ and $d(t)$, respectively.  We assume that $d(t) = g(t) + \epsilon$, where $\epsilon$ is a random variable describing the noise in measurements. In Section~\ref{EDEConsistency}, we assume that $\epsilon = 0$, so that the data are completely noise-free.     In later sections we address the case of noisy data.  
In Sections~\ref{EDEConsistency} and \ref{noisyData} we assume that the model is correct;  that is, there is some $\omega_0 \in \Omega$ so that $f(\omega_0,t)=g(t)$ for all $t \in [0,T]$.  We relax this assumption in Section~\ref{modelMismatch}.

\subsection{Expected Dynamics Estimator (EDE)}

A given data set $(t_1,d_1), \ldots, \linebreak (t_n,d_n)$ will be used to induce a probability distribution on the parameter space.  We do this through the normalized likelihood function, 
\[
		p_n(\omega)=c_n \exp(-\sum_{i=1}^n{(d_i - f(\omega, t_i))^2}),
\]
(or a variant of this expression), where $c_n$ is a constant so that $p_n$ is a probability distribution on $\Omega$. (Note that if no data has been observed, the distribution $p_0$ is just the uniform distribution in $\Omega$.)

The expected dynamics estimator (EDE) with respect to this probability distribution is then
\[
\hat{D}_n(t)=E_{p_n(\omega)}[f(\omega,t)],
\]
which we use as an estimator of the system's true dynamics.  Thus, instead of trying to maximize the likelihood function in order to estimate dynamics, we average the output dynamics, weighted by the probability as determined by the likelihood function.  It is also worth noting that the EDE is the natural estimator that is used frequently as a part of the ensemble method, and is usually computed by Monte Carlo Markov Chain methods. 

\section{EDE Consistency for noise-free data}
\label{EDEConsistency}
In this section, we establish results about the consistency of the expected dynamics estimator, that is, the ability to recover the true dynamics under a specified experimental design. The proof will be provided for two different cases:
\begin{enumerate}
	\item When the sampled time points $\{t_n\}$ are chosen at random from an absolutely continuous probability distribution $\mu$ on [0,T].
	\item When the sampled time points are chosen sequentially as in \cite{M} and \cite{D}, where the next sampled time point will be the point with highest current uncertainty (output variance).
\end{enumerate}

Before moving forward to analyze the convergence of the EDE in these two cases, it is worth mentioning the distinction between two different sources of uncertainties (in both parameters and output dynamics):  noise in data(aleatoric uncertainty), and structural uncertainty (epistemic uncertainty) in the model.  Given a set of noise-free data, the corresponding set of parameter values that fit the data perfectly well can still be an infinite set (usually, is a union of low-dimensional manifolds).  The simplest example for this phenomenon is when the number of data is less than the number of model parameters. 

In unidentifiable nonlinear systems, this set of "fitted" parameters may not collapse to a point mass even if all measurable outputs are known completely. This uncertainty in parameters may never be eliminated. The forward propagation of this uncertainty to the output space is the target in this noise-free framework.

The likelihood proposed in the noise-free setting, therefore, is not associated with noise in data, but with the structural uncertainty in model parameters from available data (how well a parameter set fits the data). Instead of focusing on a low-dimensional set of "fitted" parameters, we use an everywhere positive likelihood function to constrain the parameter space. From a methodological point of view, the idea here is similar to those behind simulated annealing methods for optimization and multiple Monte Carlo Markov Chains method for statistical inference: since the objects of interest is difficult to identify, we relax it by heated objects that are easier to study and use our experimental design algorithm to sequentially reduce the temperature in an optimal way to identify the true output dynamics.

\subsection{Randomly chosen experimental design points}
To illustrate the ideas used in later results, we consider the case when the sampled time points $\{t_n\}$ are chosen independently at random from an absolutely continuous probability distribution, with the assumption that the data are noise free (i.e. $d(t_i)=g(t_i)$ for all $i$).  In this setting, we have the following theorem, which says that in the limit when $n \to \infty$, the expected dynamics estimator converges to the system's true dynamics.
\begin{theorem}  Suppose there exists $\omega_0 \in \Omega$ such that $f(\omega_0, t) = g(t)$ for all $t \in [0,T]$.  Suppose also $\{t_n\}$ are chosen independently at random from an absolutely continuous probability distribution $\mu$ on [0,T] and that $1 \le r < \infty$. Let
\[
p_n(\omega)=c_n ~ \exp\left(-\sum_{i=1}^n{\left|f(\omega,t_i)-g(t_i)\right|^r}\right),
\]
where $c_n$ is the normalizing constant to ensure that $p_n$ is a probability distribution on $\Omega$.  Then for all $t \in [0,T]$,
\[
\lim_{n \to \infty}{E_{p_n}[f(\omega,t)]}=g(t).
\]
Moreover, the convergence is uniform in $t$.
\label{Theo1}
\end{theorem}

Before proving the theorem, we provide some intuition.  Every time a new time point is sampled, the likelihood function is multiplied by a new term of the form $\exp\left({-\left|f(\omega,t_{n+1})-g(t_{n+1})\right|^r}\right)$.
If $\omega$ does not correspond to the true dynamics, there must be a region of $[0,T]$ where $f(\omega,t)$ differs from $g(t)$.  Since the $\{t_n\}$ are chosen independently at random from an absolutely continuous probability distribution, eventually multiple time points will be sampled in this region, causing the value of the likelihood at $\omega$ go to 0. Therefore, in the limit when $n \to \infty$, the distribution $p_n(\omega)$ will  concentrate more and more on the set of $\omega$ which corresponds to the true dynamics. Hence the expected dynamics will also converge to the system's true dynamics.

We use the following two lemmas, whose proofs will be provided in Section~\ref{proveLemmas}.  The first is a result on the convergence of Monte Carlo integration.  The second is a result on the convergence of the EDE. 

\begin{lemma} 
Let points $t_i$ be chosen as in Theorem~\ref{Theo1}, and let $1 \le r < \infty$.  Define
\[
h_n(\omega)=\exp\left(-\frac{1}{n}\sum_{i=1}^n{\left|f(\omega,t_i)-g(t_i)\right|^r}\right)
\]
and 
\[
h(\omega)=\exp\left(-\int_0^T{\left|f(\omega,t)-g(t)\right|^r}d\mu(t) \right).
\]
Then
	\[ 
	\frac{h_n(\omega)}{h(\omega)} \rightarrow  1 \h  \text{ uniformly in } \omega \in \Omega
	\]
and
	\[
	\lim_{n\to \infty}{\left\|h_n\right\|_n} = \left\|h\right\|_{\infty}.
	\]
\label{LemRan}
\end{lemma}

\begin{lemma} 
Let $a$ and $b$ be continuous functions on $\Omega \times [0,T]$ and $[0,T]$, respectively, and let $\{p_n\}$ be a sequence of probability distributions on $\Omega$.

\medskip
a) Define
\[
h(\omega)=\exp\left(-\int_0^T{\left|a(\omega,t)-b(t)\right|^r}d\mu(t) \right),
\]
and suppose that 
\begin{itemize}
	\item[(i)] for any $\alpha < 1$, there exists $\delta < 1$ and $C>0$ such that if $\omega \in \Omega$ with $h(\omega) \leq \alpha \left\|h\right\|_{\infty}$,  then 
   $p_n(\omega) < C \delta^n ~\forall n$;

	\item[(ii)] there exists $\omega_0 \in \Omega$ such that $a(\omega_0,t)=b(t)$ for all $t \in [0,T]$.
\end{itemize}
Then 
\[
\lim_{n \to \infty}{E_{p_n}\left[a(\omega,t)\right]}=b(t)\h \forall t \in  [0,T]
\]
and
\[
\lim_{n \to \infty}{\var_{p_n}\left[a(\omega,t)\right]}=0\h \forall t \in  [0,T].
\]
Moreover, for both limits, the convergence is uniform in $t$.

\medskip
b)  Assume that $\Omega$ is finite and that there exists a set $S \subset [0,T]$ such that
\[
\{\omega \in \Omega: p_n(\omega) \not\to 0\} \subset \{\omega \in \Omega: a(w,t)=b(t) ~ \forall t\in S\}.
\]
Then for all $t$ in $S$, 
\[
\lim_{n \to \infty}{E_{p_n}\left[a(\omega,t)\right]}=b(t)
\]
and
\[
\lim_{n \to \infty}{\var_{p_n}\left[a(\omega,t)\right]}=0.
\]
Moreover, for both limits, the convergence is uniform in $t$.

\label{Lem2}
\end{lemma}

\begin{proof}[Proof of Theorem \ref{Theo1}]
Let
\[
q_n(\omega)=\exp\left(-\sum_{i=1}^n{\left|f(\omega,t_i)-g(t_i)\right|^r}\right).
\]
Then $p_n = c_n q_n$, and $q_n = (h_n)^n$, where $h_n$ is defined as in Lemma \ref{LemRan}, so 
\[
p_n(\omega)=\frac{q_n(\omega)}{\int_{\Omega}{q_n(\omega)d\omega}}=\frac{h_n^n(\omega)}{\int_{\Omega}{h_n^n(\omega)d\omega}}=\left(\frac{h_n(\omega)}{\left\|h_n\right\|_n}\right)^n. 
\]

Let $0 < \alpha<1$, and suppose $\omega\in \Omega$ with $h(\omega)\le \alpha \left\|h\right\|_{\infty}$.  By Lemma \ref{LemRan} we have $\lim_{n \to \infty}{h_n(\omega)}= h(\omega)$ and $\lim_{n \to \infty}{\left\|h_n\right\|_n} = \left\|h\right\|_{\infty}$. Let  $\epsilon > 0$ and $\delta = \alpha(1+\epsilon)^2$.  For $\epsilon$ small, we have $\delta < 1$, and for this $\epsilon$ there exists $N$ (independent of $\omega$) large enough such that if $n > N$, then 
\[
h_n(\omega) \le (1 + \epsilon)h(w) \le \alpha(1+\epsilon) \|h\|_\infty \le  \alpha(1+\epsilon)^2 \left\|h_n\right\|_{n}.
\]
Hence for all $n > N$, 
\[
p_n(\omega)=\left(\frac{h_n(\omega)}{\left\|h_n\right\|_n}\right)^n \le \delta^n, 
\]
with $\delta < 1$.  Since there exists $\omega_0 \in \Omega$ such that $f(\omega_0,t)=g(t)$ for all $t\in [0,T]$, we can apply Lemma \ref{Lem2} (a) with $a = f$ and $b = g$ to obtain the uniform convergence
\[
\lim_{n \to \infty}{\int_{\Omega}{p_n(\omega)f(\omega,t)~d\omega}}=g(t)\h \text{for all } t \in [0,T].
\]
The integral on the left is $E_{p_n}[f(\omega,t)]$, so this gives the desired equality. 
\end{proof}

Note that the proof depends on the sequence $\{t_i\}$ only through Lemma~\ref{LemRan}, so the result holds for any sequence that yields the conclusion in that lemma.  A quasi-random sequence satisfying a low-discrepancy condition \cite{morokoff} is one such sequence, so we make the following remark.

\begin{R}
The conclusion of Theorem~\ref{Theo1} is still valid if $\{t_i\}$ is a low-dis\-crep\-an\-cy sequence, i.e.
\[
D_N(\{t_1,...,t_N\})\mathrel{\mathop:}= \sup_{B \subset \Omega}{\left|\frac{\#\{1\le i \le N: t_i \in B\}}{N}-\text{Vol}(B)\right|} \to 0
\]
when $N$ approaches infinity.
\end{R}

The results in this section imply that if data is collected uniformly at random, we can recover the true dynamics from the sampled data.  This is one example of a so-called space-filling design \cite{cox}.  However, in practice, randomly chosen points do not produce an efficient experimental design, since many of the measurements will not give much information about the system; the convergence, although guaranteed, may be slow.

\subsection{Design Points Using the Maximally Informative Next Experiment}

Intuitively, we expect the sequential designs of \cite{M} and \cite{D}, for which the next sampled time point is the one that has the highest current uncertainty (variance) to increase the convergence rate relative to randomly selected design points.  On the other hand, the measured points may no longer be dense in $[0,T]$, so it's not clear that the dynamics may be recovered on the entire interval.  

In the following theorem, we extend the consistency result in the previous subsection to this type of sequential design, with the additional assumption that $\Omega$ is finite. This assumption was also used in the context of parameter identification in \cite{LP2009} and \cite{LP2010}. We conclude that we can recover the entire true dynamics, even if all the measurements are made in a small subset of $[0,T]$ (in the extreme case, at one point).  As in the previous subsection, we still assume that data are subject to no error.

\begin{theorem}
Let $\omega_0$, $r$, $p_n$ be as in Theorem \ref{Theo1} and assume that $\Omega$ has finite cardinality.  Suppose that each $t_{n+1}$ is chosen so that
\begin{equation}
\var_{p_n(\omega)}\left[f(\omega,t)\right] \le \var_{p_n(\omega)}\left[f(\omega,t_{n+1})\right] \h \forall t \in [0,T].
\label{equa1}
\end{equation}
Then
\[
\lim_{n \to \infty}{E_{p_n}[f(\omega,t)]}=g(t) \h \forall t \in [0,T].
\]
That is, the EDE converges to the true dynamics of the system. Moreover, the convergence is uniform in $t$.
\label{Theo2}
\end{theorem}


By choosing the next time point to be the point with highest variance, we put a constraint on the variance of the whole dynamics: variance at other points must be smaller than variance at the measured points, which in turn converges to 0.  In this case we deduce that the expected dynamics on the whole interval converges to some limit dynamics. If we can prove further that a ``true" parameter vector $\omega_0$ is still in the support of the limit distribution, then obviously this limit dynamics is equal to the true system dynamics. 

As above, this is straightforward when the $t_i$ are chosen at random from an absolutely continuous distribution. However, the case when $\Omega$ is an open set and $t_i$ are chosen according to $\eqref{equa1}$ is a bit different. In a continuous framework, a parameter vector has measure zero and good performance of the true parameter vector does not guarantee that it will stay in the support of the limit distribution. Such a situation can happen in the case when the model is not robust around the true parameter and at the chosen time points, the neighbourhood around true parameters in the parameter space fit the data worse than some other regions. This may cause the expected dynamics to converge to incorrect dynamics.  Though this situation is perhaps unlikely to happen in practice, we cannot exclude such a possibility for a convergence result. 

To resolve this issue, we assume in Theorem \ref{Theo2} that $\Omega$ is a finite set. This may be achieved, for example, by subdividing each coordinate axis using a fixed step size and taking the set of points in $\Omega$ that lie on the resulting grid. An alternative approach in which the outputs and the set of possible measured time points are discretized instead of $\Omega$ is also suggested in Theorem~\ref{ThmDiscrete}. Both assumptions are natural and do not hinder the applicability of the method in practice.
 
\medskip
\begin{proof}[Proof of Theorem \ref{Theo2}]
As in the proof of Theorem \ref{Theo1}, let
\[
q_n(\omega)=\exp\left(-\sum_{i=1}^n{\left|{f(\omega,t_i)-g(t_i)}\right|^r}\right),
\]
and recall that $p_n = c_n q_n$.  Also, let $A$ be the set of cluster points of $\{t_n\}$:  points $t \in [0,T]$ such that there exists a subsequence $\{t_{n_k}\}$ of with $t_{n_k} \to t$.  

We claim first that if $p_n(w)$ does not tend to 0 with $n$ (so that $\omega$ has probability above some fixed $\rho>0$ for infinitely many $n$), then $f(\omega, t) = g(t)$ for all $t \in A$.  
Indeed, consider any $\omega \in \Omega$, $t \in A$ such that $|f(\omega,t)-g(t)|=c >0$.  Since $A$ is the set of limit points of $\{t_n\}$, there exists a subsequence $\{t_{n_k}\}$ of $\{t_{n}\}$ such that $t_{n_k} \to t$. Since $f$ and $g$ are continuous, for $k$ large enough, we have $|f(\omega,t_{n_k})-g(t_{n_k})|\ge c/2$.
Hence
\[
\sum_{i=1}^n{\left|{f(\omega,t_i)-g(t_i)}\right|^r} \to \infty
\]when $n \to \infty$, and so $q_n(\omega) \to 0$.

On the other hand, the assumption that some $\omega_0$ gives the true dynamics implies that $f(\omega_0, t)-g(t) = 0$ for all $t$, hence $q_n(\omega_0) = 1$.
Therefore, $p_n(\omega_0)/p_n(\omega) \to \infty$.  Since $\Omega$ is a finite space, $p_n(\omega_0) \le 1$, and hence $p_n(\omega) \to 0$.
Hence $p_n(w) \not \to 0$ implies $f(\omega,t)=g(t)$  for all $t \in A$.  

Using Lemma \ref{Lem2} (b) (for finite $\Omega$) with $a=f$ and $b=g$, we deduce that
\[
\lim_{n \to \infty}{E_{p_n}[f(\omega,t)]}=g(t)\h \forall t \in A
\]
and 
\[
\var_{p_n(\omega)}\left[f(\omega,t)\right] \to 0 \h \forall t \in A.
\]
On the other hand, the choice of $t_{n+1}$ gives
\begin{equation} \label{varLT}
\var_{p_n(\omega)}\left[f(\omega,t)\right] \le \var_{p_n(\omega)}\left[f(\omega,t_{n+1})\right] \h \forall t \in [0,T].
\end{equation} 

Now we claim that
\begin{equation}  \label{var0}
\var_{p_n(\omega)}\left[f(\omega,t_{n+1})\right] \to 0.
\end{equation}

Indeed, by contradiction, assume that there exists a subsequence $\{t_{n_k}\}$ and a positive constant $C$ such that
\[
\var_{p_{n_k}(\omega)}\left[f(\omega,t_{n_k+1})\right] \ge C
\]
for all $k$.  Since $[0,T]$ is compact, we can drop to a subsequence to assume that $t_{n_{k}+1}$ converges to some $t_0 \in A$.
By the continuity of $f$ and its derivatives on the compact set $\Omega \times [0,T]$, there is $C_0 > 0$ so that for all $k>0$ and $\omega \in \Omega$,
\begin{equation} \label{eq:fReg}
\left|f(\omega,t_{n_k})-f(\omega,t_0)\right| \le C_0 \left|t_{n_k}-t_0\right|.
\end{equation}
Hence by using this inequality, we have
\[
\lim \sup_{k \to \infty}{E_{p_{n_k}(\omega)}\left|f(\omega,t_{{n_k}+1})-f(\omega,t_{0})\right|} \le \lim_{k \to \infty}{C_0 \left|t_{n_{k+1}}-t_0\right|}=0
\]
which implies that
\[
\lim_{k \to \infty}{E_{p_{n_k}(\omega)}\left[f(\omega,t_{{n_k}+1})\right]}=\lim_{k \to \infty}{E_{p_{n_k}(\omega)}\left[f(\omega,t_{0})\right]}.
\]
By a similar argument, we also have
\[
\lim_{k \to \infty}{E_{p_{n_k}(\omega)}\left[f^2(\omega,t_{{n_k}+1})\right]}=\lim_{k \to \infty}{E_{p_{n_k}(\omega)}\left[f^2(\omega,t_{0})\right]}.
\]
Therefore
\[
\lim_{k \to \infty}{\var_{p_{n_k}(\omega)}\left[f(\omega,t_{{n_k}+1})\right]} = \lim_{k \to \infty}{\var_{p_{n_k}(\omega)}\left[f(\omega,t_{0})\right]} =0,
\]
which contradicts the choice of $C$.

From \eqref{varLT} and \eqref{var0} we obtain
\[
\var_{p_n(\omega)}\left[f(\omega,t)\right] \le \var_{p_n(\omega)}\left[f(\omega,t_{n+1})\right] \to 0 \h \forall t \in [0,T].  
\]
In other words, for all $t$ in $[0,T]$,
\begin{equation} \label{limvar}
\lim_{n \to \infty} \sum_{\omega \in \Omega}{p_n(\omega)\left(f(\omega,t)-E_{p_n}\left[f(\omega,t)\right]\right)^2} = 0.
\end{equation}

The fact that $\omega_0$ gives the true dynamics implies that $\omega_0$ is a maximum for $q_n$, hence for $p_n$.  Hence $p_n(\omega_0) \ge p_n(\omega)$  for all $\omega \in \Omega$, and from the fact that $\Omega$ is finite, we deduce that $p_n(\omega_0) \ge 1/|\Omega|$.  Using this with \eqref{limvar} gives
\begin{align*}
\left(f(\omega_0,t)-E_{p_n}\left[f(\omega,t)\right]\right)^2 &\le  |\Omega|p_n(\omega_0) \left(f(\omega_0,t)-E_{p_n}\left[f(\omega,t)\right]\right)^2
\\
&\le |\Omega| \sum_{\omega \in \Omega}{p_n(\omega) \left(f(\omega,t) - E_{p_n}\left[f(\omega,t)\right]\right)^2}\\
&\to 0,
\end{align*}
as $n \to \infty$.  Hence
\[
E_{p_n(\omega)}\left[f(\omega,t)\right] \to f(\omega_0,t)=g(t) \h \forall t \in [0,T].
\]
\end{proof}

As in the discussion before the proof, the only reason we use the variance criterion is to bound the variance of dynamics at unmeasured points by the variance at measured points. This criterion can be relaxed as follows.

\begin{theorem}
The result from Theorem \ref{Theo2} is still valid if Condition \eqref{equa1} (that the next time point has maximum variance) is replaced by the condition that the variance at the next time point is within a fixed constant of the maximum variance.  That is, there exists $C>1$ so that for all $t \in [0,T]$, 
\begin{equation}
\var_{p_n(\omega)}\left[f(\omega,t)\right] \le C ~\var_{p_n(\omega)}\left[f(\omega,t_{n+1})\right].
\label{equa2}
\end{equation}
\label{ThmWithC}
\end{theorem}

There are several motivations for this relaxation of criterion \eqref{equa1}. First, in practice, the optimization of the variance function (which is usually done by Markov Chain Monte Carlo methods) is subject to random effects arising in the sampling process. By using criterion \eqref{equa2}, we look for a near-optimal solution of the optimization problem; this condition is stable with respect to a MCMC scheme. Second, in real experiments, some sets of measurements may be more expensive and technically difficult than the others. By looking for a near-optimal solution, we make it possible for experimenters to find an alternative measurement when the optimization problem gives rise to an optimum that is experimentally difficult to implement. Finally, as we will see in Section~\ref{noisyData}, the new criterion allows us to apply additional criteria for point selection in order to facilitate resampling, which will be used to establish convergence rate of the EDE in the face of noisy data. 

As noted above, instead of discretizing $\Omega$, we may discretize the output space and the measurement space.  This gives the following result. 

\begin{theorem}
The result from Theorem \ref{Theo2} is still valid if we assume that $\Omega$ is open and bounded, but that the possible outputs of the system and the set of possible measured time points are both finite.  In this case, we get convergence of the EDE on the full (finite) set of possible measured time points. 
\label{ThmDiscrete}
\end{theorem}

\begin{proof}
We denote by $\T=\{\tau_i\}_{i=1}^K$ the set of all possible measured time points and assume that at each time point, the output function $f(\omega,t)$, as well as the true dynamics $g(t)$, are discretized by a finite grid. That is, the continuous function $g(t)$ is approximated by values $(Rg(\tau_1), \ldots, Rg(\tau_K))$, where $Rg(t)$ is obtained from $g(t)$ by rounding to the nearest of some finite set of allowable output values. 

By making arbitrarily small perturbations to the possible output values if necessary, we can assume without loss of generality that $\forall \tau \in \T$, the true output value $g(\tau)$ does not lie midway between two allowable output values. Therefore, there exists an open neighbourhood $U_{\omega_0}$ of $\omega_0$  such that if $\omega \in U_{\omega_0}$, then $Rf(\omega,\tau)=Rf(\omega_0,\tau)=Rg(\tau) ~~ \forall \tau \in \T$.  For the remainder of the proof we use $f$ and $g$ to mean $Rf$ and $Rg$.  

As in the proof of Theorem \ref{Theo2}, we now consider any $\omega \in \Omega$ such that $p_n(\omega)$ does not tend to 0 with $n$. Assume that $f(\omega,t) \ne f(\omega_0,t)$ for some $t$ in the cluster set, $A$, of $\{t_i\}$ (since $\T$ is finite here, this is the set of points that are measured infinitely many times). Using the same argument as in the proof of Theorem \ref{Theo2}, we deduce that $p_n(\omega) \le p_n(\omega_0)$ and $p_n(\omega_0)/p_n(\omega) \to \infty$. Note that in this case, although $\Omega$ is not finite, the argument is still valid since $p_n$ is constant on the open set $U_{\omega_0}$, so that $p_n(\omega_0)$ is bounded above by $1/\V(U_{\omega_0})$, where $\V(U)$ denotes the volume of a set $U$.

Therefore, $p_n(\omega) \not \to 0$ implies $f(\omega,t)=g(t)$  for all $t \in A$. Since $\Omega$ may be infinite, Lemma \ref{Lem2} cannot be applied directly in this case. However, by denoting $U_A=\{\omega \in \Omega: f(\omega,t)=g(t)  ~\forall t \in A\}$, we have for all $t \in A$
$$
\left|\int_{\Omega}{p_n(\omega)f(w,t)d\omega}-g(t) \right| \le 
\int_{\Omega \setminus {U_A}} {p_n(\omega) \left|f(w,t)-g(t)\right|  \,d\omega}.
$$
Since $f$ and $g$ are bounded and $p_n(\omega_0) < 1/\text{Volume}(U_{\omega_0})$, we have $|p_n(\omega)|f(w,t)-g(t)| \le C p_n(\omega_0)\le C/\V(U_{\omega_0})$.  Also, $p_n(w) \to 0$ on $\Omega \setminus {U_A}$, so by the Dominated Convergence Theorem, the right hand side converges to 0 as $n$ tends to $\infty$. We deduce that $E_{p_n(\omega)}[f(\omega,t)] \to g(t) ~~ \forall t \in A$. By a similar argument, we also have $\var_{p_n(\omega)}[f(\omega,t)] \to 0 ~~ \forall t \in A$.

We next use the fact that the set $\T=\{\tau_i\}_{i=1}^K$ of possible measured time points is finite to deduce that $\var_{p_n(\omega)}[f(\omega,t_{n+1})] \to 0$. Indeed, assume that 
$$
\var_{p_{n_k}(\omega)}[f(\omega,t_{n_k+1})] \ge C
$$ 
for some subsequence $\{n_k\}$ and positive constant C. Since $\T$ is finite, there exists $t_0 \in A$ that appears in the subsequence $\{t_{n_k}\}$ infinitely many times; this implies that $\var_{p_{n_k}(\omega)}[f(\omega,t_0)] \not \to 0$, which is a contradiction. 

Hence, $\var_{p_n(\omega)}[f(\omega,t_{n+1})] \to 0$. Combining this with \eqref{equa1}, we see that in fact $\var_{p_n(\omega)}[f(\omega,t)] \to 0$ for all $t \in \T$.  This proves that the EDE  converges to the true system dynamics on $\T$. 
\end{proof}

Note that the condition of discrete measured time points in the previous result allows us to avoid the need for the regularity condition \eqref{eq:fReg}, which does not hold for the piecewise constant functions obtained by discretizing the system outputs.  In the next section we provide further justification for a finite set of measurement points.

\section{EDE Consistency with Noisy Data} \label{noisyData}
In practice, of course, data from experiments are subject to noise.  Hence
in this section we extend the results from previous sections to the case of additive Gaussian noise.  As is common in many settings, we assume that
\[
d(t_i)=g(t_i)+\epsilon_i
\]
where $g(t)$ is the true dynamics (which is unknown), $d(t_i)$ is the measured data at the sampled time point $t_i$, and $\epsilon_i$ are i.i.d. Gaussian random variables (see \cite{D} for emperical support of this noise model).

The analysis in the case of noisy data is a bit different from that used in the previous section. Intuitively, if "close", but not exactly the same points in time, $t_1$ and at $t_2$, are measured, and if there is a functional relation between the output at $t_1$ and at $t_2$, then the measurement at $t_1$ will also help refine the information about data at $t_2$. However, theoretically, this assertion is difficult to prove and may even be incorrect, due to nonlinearity: if the relation between output at $t_1$ and $t_2$ are nonlinear, using data at $t_2$ to constrain $t_1$ may create a bias in the fitted output. 

For example, if $f_2 =f_1^2$ then
\[(f_1+e)^2=f_1^2+2ef_1+e^2=f_2 + e f_1+e^2\]
When using averaging, the linear error term will go away by the strong law of large number, but the quadratic term will have positive expectation, which results in a bias in estimation of $f_2$.  The stronger the non-linearity is, the larger the bias and that makes it hard clarify the convergence.

In order to obtain a convergence result using noisy data, we need to be able to average over multiple trials, which makes sense only if we measure repeatedly at a given time.  In the theorem below, as in Theorem~\ref{ThmDiscrete}, we discretize the time interval and allow measurements to be taken at only finitely many specified points and use a slightly different form of probability distribution.  This guarantees that experiments will be replicated many times at ``important" points.  When data are collected multiple times, the average value is used to constrain the dynamics: the larger the number of times we make the measurement at a time point $t$, the more confidence we put on the average data at that point. With this framework, we again have the convergence of the EDE to the true system dynamics.


The idea of using a finite grid to replace the whole time interval to facilitate resampling is a common technique in the problem of parameter identification (\cite{LP2009},\cite{LP2010}). In studies of ODEs, under the assumption that $f$ is analytic, this is further supported by the following theorem from \cite{Sontag02fordifferential}, which guarantees that if we can identify the system dynamics on a finite grid, we can identify the dynamics on the whole interval.
\begin{theorem}(Sontag \cite{Sontag02fordifferential})

Assume $f(w, t)$ depends analytically on $\omega$ and $t$, and let $N$ be the dimension of the parameter space. Then, for Lebesgue almost every randomly chosen set of $2N+1$ experiments, the following property holds: For any two parameters that have distinct dynamics, one of the experiments in this set will distinguish them.
\label{Sontag}
\end{theorem} 

We further note that the assumption of analyticity may be replaced by an assumption that $\Omega$ is a finite set.  That is, with this assumption we can find a finite grid $\T \subset [0,T]$ that satisfies the above property: For any two parameters that have distinct dynamics on $[0, T]$, one of the experiments $t \in \T$ in this grid will distinguish them.

Finally, even in the case when the parameter space $\Omega$ is an open set, by choosing the discretized time points to be the nodes for an efficient interpolation scheme, then by interpolating on this finite set, convergence on the finite set of times $\T$ converts to uniform approximation on the entire interval $[0,T]$.

Hence throughout this section, we discretize $[0,T]$ to a finite grid $\T=\{\tau_i\}_{i=1}^K$, and assume that the experiments can be made only at the nodes of this grid.  We also continue to assume that $\Omega$ has finite cardinality.  With these assumptions, we have the following theorem. 
 
\begin{theorem}
Let $C>1$. Assume that $\Omega$ is finite and at step $n$, $t_{n+1} \in \T$ is chosen so that
\[
\var_{p_n(\omega)}\left[f(\omega,t)\right] \le C~\var_{p_n(\omega)}\left[f(\omega,t_{n+1})\right] \h \forall t \in \T.
\]
For $1 \le k \le K$, let $k_n(\tau_i)$ be the number of experiments made at time $\tau_i$ up through step $n$ and $\{d_j(\tau_i)\}_{j=1}^{k_n(\tau_i)}$ be the data values from those experiments, with $d_j(\tau_i) = g(\tau_i) + \epsilon$; the $\epsilon$ are iid $N(0, \sigma^2)$.  Define $B_n=\{\tau_i: k_n(\tau_i)>0\}$ and
\[
p_n(\omega)=c_n ~ \exp\left(- \sum_{\tau_i \in B_n}{k_n(\tau_i){\left({f(\omega,\tau_i)-\frac{1}{k_n(\tau_i)}\sum_{j=1}^{k_n(\tau_i)}{d_j(\tau_i)}}\right)^{r}}}\right),
\]
where $c_n$ is the normalizing constant { and $r>2$ }.  Then
\[
\lim_{n \to \infty}{E_{p_n}\left[f(\omega,t)\right]} =g(t)\h \forall t \in \T.
\] 
Moreover, the convergence is uniform in $t \in \T$.

{
The same result with $r=2$ is also valid if the following condition is satisfied
\begin{equation}
\lim_{n\to \infty}{\frac{\log \log k_n(\tau_1)}{k_n(\tau_2)}} =0 \h \forall \tau_1, \tau_2 \in A
\label{e9}
\end{equation}
where $A$ is the set of all cluster points of $\{t_n\}$.
\label{Theo5}
}
\end{theorem}

Note that if $\T$ satisfies the conditions in the discussion following Theorem~\ref{Sontag}, then determining the dynamics in $\T$ is sufficient to determine the dynamics on all of $[0,T]$.  

\begin{proof}

Let $A=\{\tau_i: \lim_{n\to \infty}{k_n(\tau_i)} = \infty\}$
and
\[
q_n(\omega)=\exp\left(- \sum_{\tau_i \in B_n}{k_n(\tau_i){\left({f(\omega,\tau_i)-\frac{1}{k_n(\tau_i)}\sum_{j=1}^{k_n(\tau_i)}{d_j(\tau_i)}}\right)^{ r}}}\right).
\]

We claim that
\[
\{\omega: q_n(w) \not \to 0\} \subset \{\omega: f(\tau_i,\omega)=g(\tau_i), \forall \tau_i \in A\}.
\]

{

Indeed, consider any $\omega \in \Omega$, $\tau_{i_0} \in A$ such that $|f(\omega,\tau_{i_0} )-g(\tau_{i_0} )|=c>0$. Let  $X_j=d_j(\tau_{i_0} )=g(\tau_{i_0} )+\epsilon$ and note that $\{X_j\}$ is a Gaussian sequence of iid random variables with $E[X_j]=g(\tau_{i_0} )=f(w_0,\tau_{i_0} )$.  By the law of large numbers, we have with probability 1
\[
g(\tau_i)= \lim_{n \to \infty}{\frac{1}{k_n(\tau_i)}\sum_{j=1}^{k_n(\tau_i)}{d_j(\tau_i)}}
\]
Hence there exists $N$ such that for all $n>N$
\[
\left|g(\tau_{i_0} )- \lim_{n \to \infty}{\frac{1}{k_n(\tau_{i_0} )}\sum_{j=1}^{k_n(\tau_{i_0} )}{d_j(\tau_{i_0} )}}\right| \le c/2 
\]
which implies (by triangle inequality)
\[
\left|f(\tau_{i_0} ,\omega)- \lim_{n \to \infty}{\frac{1}{k_n(\tau_{i_0} )}\sum_{jl=1}^{k_n(\tau_{i_0} )}{d_j(\tau_{i_0} )}}\right| \ge c/2
\]
Therefore
\begin{equation}
\sum_{\tau_i \in B_n}{k_n(\tau_i){\left({f(\omega,\tau_i)-\frac{1}{k_n(\tau_i)}\sum_{j=1}^{k_n(\tau_i)}{d_j(\tau_i )}}\right)^r}} \ge (c/2)^r k_n(\tau_{i_0} ) \to \infty
\label{e7}
\end{equation}
as $n \to \infty$. 

\bigskip

Now consider any $\tau_i \in A$.  By the law of the iterated logarithm, we have with probability 1
\[
\lim \sup_{n \to \infty}{\sqrt{\frac{k_n(\tau_i)}{\log \log k_n(\tau_i)}}\left({f(\omega_0,\tau_i)-\frac{1}{k_n(\tau_i)}\sum_{j=1}^{k_n(\tau_i)}{d_j(\tau_i )}}\right)}=\sqrt{2}
\]
If $r>2$, there exists a constant $C$ such that
\begin{equation}
k_n(\tau_i)\left({f(\omega_0,\tau_i)-\frac{1}{k}\sum_{j=1}^{k}{d_j(\tau_i)}}\right)^{ r } \le C \frac{\log \log k_n(\tau_i)}{(k_n(\tau_i))^{(r/2-1)}} \to 0
\label{e8}
\end{equation}
as $n \to \infty$.

Since this is true for any $\tau_i$ in the finite set $A$, we have
\[
\lim_{n \to \infty}{\sum_{\tau_i \in B_n}{k_n(\tau_i){\left({f(\omega_0,\tau_i)-\frac{1}{k_n(\tau_i)}\sum_{j=1}^{k_n(\tau_i)}{d_j(\tau_i)}}\right)^{ r}}}} < \infty.
\]
Therefore, $q_n(\omega_0)$ is bounded below, so $\frac{p_n(\omega_0)}{p_n(\omega)} = \frac{q_n(\omega_0)}{q_n(\omega)} \to \infty$. Since $\Omega$ is a finite space, $p_n(\omega_0) \le 1$. This makes $p_n(\omega) \to 0$.
}
{

In the case when $r=2$, we have
\[
k_n(\tau_i)\left({f(\omega_0,\tau_i)-\frac{1}{k}\sum_{j=1}^{k}{d_j(\tau_i)}}\right)^2 \le C\log \log k_n(\tau_i)
\]
which implies
\[
\lim \sup_{n \to \infty}{\sum_{\tau_i \in B_n}{k_n(\tau_i){\left({f(\omega_0,\tau_i)-\frac{1}{k_n(\tau_i)}\sum_{j=1}^{k_n(\tau_i)}{d_j(\tau_i)}}\right)^2}}} \le C_1 + \sum_{\tau_i \in A}{C\log \log k_n(\tau_i)}.
\]
Equation $\eqref{e9}$ implies that there exists $N$ such that for all $n \ge N$
\[
\log \log k_n(\tau_i) \le \frac{(c/2)^2}{2C(\#A)} k_n(\tau_{i_0}) \h \forall \tau_i \in A
\]
where $\# A$ denotes the cardinality of $A$, and $c$ is the constant in $\eqref{e7}$.

Combine this inequality and $\eqref{e7}$ (with $r=2$), we have
\[
\frac{p_n(\omega)}{p_n(\omega_0)} = \frac{q_n(\omega)}{q_n(\omega_0)} \le \exp \left(C_1-\frac{c^2}{8} k_n(\tau_{i_0})\right) \to 0
\]
as $n \to \infty$. Hence $p_n(\omega) \to 0$.
}

At this point, we have proved that 
\[
\{\omega: p_n(w) \not \to 0\} \subset \{\omega: f(\omega,\tau_i)=g(\tau_i), \forall \tau_i \in A\}.
\]
Hence by Lemma \ref{Lem2} (b)
\[
\lim_{n \to \infty}{E_{p_n}[f(\omega,\tau_i)]}=g(\tau_i)\h \forall \tau_i \in A
\]
and
\[
\lim_{n \to \infty}{\var_{p_n}[f(\omega,\tau_i)]}=0 \h \forall \tau_i \in A.
\]
On the other hand, we have
\[
\var_{p_n(\omega)}\left[f(\omega,t)\right] \le C \var_{p_n(\omega)}\left[f(\omega,t_{n+1})\right] \h \forall t \in \T.
\]
Using the same argument as in the proof of Theorem \ref{Theo2}, we have
\[
\var_{p_n(\omega)}\left[f(\omega,t)\right] \to 0 \h \forall t \in \T 
\]
and
\[
E_{p_n(\omega)}\left[f(\omega,t)\right] \to f(\omega_0,t)=g(t) \h \forall t \in \T.
\]

\end{proof}

By an argument similar to that used in the proof of Theorem~\ref{ThmDiscrete}, we obtain the following result.

\begin{theorem}
The result of Theorem \ref{Theo5} is still valid if we replace the condition of finite cardinality of $\Omega$ with the condition of a finite set of output values and possible measurement time points. 
\end{theorem}

\section{EDE consistency with model mismatch} \label{modelMismatch}
So far we have investigated various schemes to design experiments for dynamics identification, under the assumption that the investigated model is a correct model, i.e.  there exists $\omega_0 \in \Omega$ such that $f(\omega_0,t)=g(t)$  for all $t \in [0,T]$. 
Here we relax this condition using the concept of $\epsilon$-equivalence.

\begin{definition} Let $\epsilon>0$ and suppose $g$ and $h$ are continuous on $[0,T]$.  Then $g$ and $h$ are $\epsilon$-equivalent means that $\|g-h\|_\infty < \epsilon$.
\end{definition}

To obtain the main result in this section, we also need to assume that the function outputs are discretized by a finite grid of resolution $\epsilon$, similar to the discretization in Theorem~\ref{ThmDiscrete}.  However, here we use an adaptive discretization in that it changes based on the measurements obtained so far and based on the time point. 

\begin{definition} 
Let $h$ be continuous on $[0,T]$, let $\T$ and $d_j(\tau_i)$ be as in Theorem~\ref{Theo5}, and let $\epsilon > 0$.  Define
$$ 
d_n^*(\tau_i)=\left\{ \begin{array}{ll} \frac{1}{k_n(\tau_i)}\sum_{j=1}^{k_n(\tau_i)}{d_j(\tau_i)} & {\rm if} \; k_n(\tau_i) > 0\\
0 & {\rm if} \; k_n(\tau_i) = 0
\end{array} \right.
$$
and
$$
R_n^\epsilon h(\tau_i)=d_n^*(\tau_i)+ \text{sgn}(h(\tau_i)-d_n^*(\tau_i)) \left \lfloor \frac{|h(\tau_i)-d_n^*(\tau_i)|}{\epsilon}\right\rfloor \epsilon.
$$
\end{definition} 

This choice of discretization is needed to guarantee convergence of the estimated
dynamics. With this setting, we have the following theorem, in which we use the framework of Theorem \ref{Theo5} but with the output discretization just given.  
The proof of this theorem is a combination of the techniques employed in Theorem \ref{Theo5} and Theorem~\ref{ThmDiscrete}.  Here $\lfloor x \rfloor$ denotes the largest integer less than or equal to $x$.

\begin{theorem}
Let $\Omega$, $C$, $\T$, $B_n$, $k_n$, $d_j(\tau_i)$ be as in Theorem \ref{Theo5}, $\epsilon_0>0$ and assume that there is $\omega_0 \in \Omega$ such that $f(\omega_0, t)$ and $g(t)$ are $\epsilon_0$-equivalent.  For $\epsilon > \epsilon_0$ define
\[
p_n(\omega)=c_n ~ \exp\left(- \sum_{\tau_i \in B_n}{k_n(\tau_i){\left({R_n^\epsilon f(\omega,\tau_i)-\frac{1}{k_n(\tau_i)}\sum_{j=1}^{k_n(\tau_i)}{d_j(\tau_i)}}\right)^2}}\right),
\]
where $c_n$ is the normalizing constant, and assume that at each step, the next measurement is chosen so that
\[
\var_{p_n(\omega)}\left[R_n^\epsilon f(\omega,t)\right] \le C~\var_{p_n(\omega)}\left[R_n^\epsilon f(\omega,t_{n+1})\right] \h \forall t \in \T.
\]

Then, for almost every $\epsilon > \epsilon_0$, the expected dynamics converges (uniformly in $t \in \T$) to limit dynamics that are $\epsilon$-equivalent to $g(t)$. 
\label{Theo6}
\end{theorem}

\begin{proof}
Denote by $A$ the set of all $t \in \T$ that are measured infinitely many times. By the strong law of large numbers, we have $d_n^*(\tau) \to g(\tau)$ for all $\tau \in A$.  Since $\Omega$ and $\T$ are finite, there is a full measure set of $\epsilon > \epsilon_0$ such that for all $\tau \in \T$ and $\omega \in \Omega$, the distance between $g(t)$ and $f(\omega,\tau)$ is not a multiple of $\epsilon$.  This implies that $\lim_{n \to \infty} |f(\omega, \tau) - d_n^*(\tau)|/\epsilon$ is not an integer for any $\omega \in \Omega$ and $\tau \in A$, hence that $\lim_{n \to \infty} R_n^\epsilon f(\omega, \tau)$ exists for all such $\omega$ and $\tau$.  Also, if 
$\tau \not \in A$, then $R_n^\epsilon f(\omega,\tau)$ is constant for $n$ large enough.  Hence $\lim_{n\to \infty}{R_nf(\omega,\tau)}$ exists for all  $ \omega \in \Omega, \tau \in \T$.

On the other hand, the assumption on $\omega_0$ implies that for all $\tau \in \T$
\[
|d_n^*(\tau_i)-f(\omega_0,\tau)| \le |d_n^*(\tau_i)-g(t)| + \epsilon_0.
\]
For $n$ sufficiently large, the right hand side is less than $\epsilon$ for all $\tau \in \T$.  For such $n$ we have $R_n^\epsilon f(\omega_0, \tau)=R_n^\epsilon g(\tau)$ for all $\tau$ in $\T$.

Now consider $\omega \in \Omega$ such that $\lim_{n \to \infty}{R_n^\epsilon f(\omega,\tau)} \ne \lim_{n \to \infty}{R_n^\epsilon f(\omega_0,\tau)}$ for some $\tau$ in $A$. Using the same argument as in the proof of Theorem \ref{Theo5}, we deduce that $p_n(\omega_0)/p_n(\omega) \to \infty$. Since $\Omega$ is a finite space, $p_n(\omega_0) \le 1$. This makes $p_n(\omega) \to 0$. We have proved that 
\[
\{\omega: p_n(w) \not \to 0\} \subset \{\omega: \lim_{n\to \infty}{R_nf(\omega,\tau)}=\lim_{n\to \infty}{R_ng(\tau)}, \forall \tau \in A\}.
\]
Then by Lemma \ref{Lem2} (b)
\[
\lim_{n \to \infty}{E_{p_n}[R_n^\epsilon f(\omega,\tau)]}=\lim_{n \to \infty}{{E_{p_n}[R_n^\epsilon  g(\tau)]}}=g(\tau)\h \forall \tau \in A
\]
and
\[
\lim_{n \to \infty}{\var_{p_n}[R_n^\epsilon  f(\omega,\tau)]}=0 \h \forall \tau \in A.
\]
On the other hand, we have
\[
\var_{p_n(\omega)}\left[R_n^\epsilon f(\omega,t)\right] \le C \var_{p_n(\omega)}\left[R_n^\epsilon f(\omega,t_{n+1})\right] \h \forall t \in \T.
\]
Using the same argument as in the proof of Theorem \ref{Theo2}, we have
\[
\var_{p_n(\omega)}\left[R_n^\epsilon f(\omega,t)\right] \to 0 \h \forall t \in \T 
\]
and
\[
\lim_{n \to \infty}{E_{p_n(\omega)}\left[R_n^\epsilon f(\omega,t)\right]} =\lim_{n \to \infty}{ R_n g(t)} \h \forall t \in \T.
\]
This proves that the EDE  converges to limit dynamics that are $\epsilon$-equivalent to the true system dynamics on $\T$. 
\end{proof}



\section{Proofs of Supporting Lemmas} \label{proveLemmas}
In this section, we provide the proofs of the two lemmas that have been used throughout this paper.

\subsection{Lemma~\ref{LemRan}}(Convergence of Monte Carlo integration)

\begin{proof}[Proof]

First, we note that
\[
\frac{h_n(\omega)}{h(\omega)}= \exp \left(\int_0^T{\left|f(\omega,t)-g(t)\right|^r}d\mu(t) - \frac{1}{n}\sum_{i=1}^n{|f(\omega,t_i)-g(t_i)|^r}\right).
\]
Using the Koksma-Hlawka inequality for convergence of quasi-Monte Carlo integration \cite{niederreiter}, we have
\begin{align}
\left|\int_0^T{\left|f(\omega,t)-g(t)\right|^r}d\mu(t) \right. &- \left. \frac{1}{n}\sum_{i=1}^n{|f(\omega,t_i)-g(t_i)|^r}\right|  \label{eq:KH} \\
&\le r D_n^{*} \int_0^T{\left|f(\omega,t)-g(t)\right|^{r-1} \left|\frac{\partial f}{\partial t}(\omega,t)-g'(t)\right|}d\mu(t), \nonumber
\end{align}
where $D_n^*$ is the discrepancy of the finite sequence $\{t_1,t_2,...,t_n\}$ (see \cite{morokoff} for more information about the discrepancy).

Since $f$ is a $C^1$ function on the compact set $\Omega \times [0,T]$ and $g$ is $C^1$ on $[0,T]$, there exists M independent of $\omega$ and $t$ such that
\begin{equation}
\int_0^T{\left|f(\omega,t)-g(t)\right|^{r-1} \left|\frac{\partial f}{\partial t}(\omega,t)-g'(t)\right| }d\mu(t) \le M.
\label{eq:M}
\end{equation}
Since $\mu$ is absolutely continuous with respect to Lebesgue measure, we have $D_n^* \to 0$ as $n \to \infty$.  

From \eqref{eq:KH} and \eqref{eq:M} we have for any $\omega \in \Omega$ that
$$ \left|\log \left( h_n(\omega)/h(\omega) \right) \right| \leq r M D_n^* \rightarrow 0. $$
Hence $h_n(\omega) /h(\omega) \rightarrow  1$ uniformly in $\omega \in \Omega$.  

Also, since $h_n(\omega) \le 1$ for all $\omega$ and $h(\omega_0) = 1$, we have 
	\begin{align}
	\limsup_{n\to \infty}	\left\|h_n\right\|_n &= \limsup_{n\to \infty}	\left(\int_{\Omega}{h_n^n~d\omega}\right)^{1/n} \nonumber \\
	&\le \limsup_{n\to \infty}	\V(\Omega)^{1/n} \nonumber \\
	&=1 = \left\|h\right\|_{\infty}. \label{eq:hInfty}
	\end{align}
	
To get a lower bound, let $\epsilon >0$.  Note that if $h(\omega) \geq 1 - \epsilon/2$ and $|h_n(\omega) - h(\omega)| \leq \epsilon/2$, then by the triangle inequality, we have $h_n(\omega) \ge 1-\epsilon$.  Also, since $h$ is continuous and $\|h\|_\infty = 1$, we have 
\[
C_\epsilon:=\V(\{\omega:h(\omega) \ge 1-\epsilon/2\}) > 0.
\]
Since $h_n/h$ converges uniformly on $\Omega$ to 1 and $|h| \leq 1$, there exists $N(\delta, \epsilon)$ large enough such that for all $n \ge N$ and all $\omega \in \Omega$, $|h_n(\omega)-h(\omega)| \le \epsilon/2.$
Hence
\[
\V(\{\omega:h_n(\omega) \ge 1-\epsilon\} ) \ge C_\epsilon.
\]
So
\[
\int_{\Omega}{h_n^n~d\omega} \ge \int_{\{h_n \ge 1-\epsilon\}}{h_n^n~d\omega} \ge C_\epsilon (1-\epsilon)^n 
\]
and
\[
\left(\int_{\Omega}{h_n^n~dx}\right)^{1/n} \ge  C_\epsilon^{1/n} (1-\epsilon).
\]

Taking $n \to \infty$, we deduce
	\[
	\left\|h\right\|_{\infty} \ge \limsup_{n\to \infty}{\left\|h_n\right\|_n} \ge \liminf_{n\to \infty}{\left\|h_n\right\|_n} \ge \left\|h\right\|_{\infty}-\epsilon.
	\]
\h Since $\epsilon$ was arbitrary, $\lim_{n\to \infty}{\left\|h_n\right\|_n}=\left\|h\right\|_{\infty}.$	
\end{proof}

\subsection{Lemma~\ref{Lem2}}(Convergence of the Expected Dynamics Estimator)
\begin{proof}[Proof]
We provide the proof for part (a). The proof for part (b) uses a similar
argument.

Let $\epsilon > 0$ and define
$$
U = \{\omega \in \Omega : |a(\omega,t)-b(t)| < \epsilon, \forall t \in [0,T]\}.
$$
Since $a,b$ are continuous and $a(\omega_0, t) = b(t)$ for all $t$, we see that $U$ is a neighborhood of $\omega_0$.  Then for $t \in [0,T]$, we have
\begin{align*}
\Big|\int_{\Omega}{p_n(\omega)a(w,t)d\omega}&-b(t) \Big| \le \int_{\Omega}{p_n(\omega)\left|a(w,t)-b(t)\right|~d\omega} \\
& =\int_{\Omega \setminus U} {p_n(\omega) \left|a(w,t)-b(t)\right|  \,d\omega} + \int_{U}{p_n(\omega) \left|a(w,t)-b(t)\right| \,d\omega}\\
& \le \int_{\Omega \setminus U}{p_n(\omega) \left|a(w,t)-b(t)\right| \,d\omega} + \epsilon
\end{align*}
Now we claim that there exists $\alpha <1$ such that $\forall \omega \in \Omega \setminus U$, $h(\omega) \le \alpha$.  Indeed, assume that $\exists \omega_n \in \Omega \setminus U$ with $h(\omega_n) \to 1$.  Then for each $n$ there is $t_n \in [0,T]$ such that $|a(\omega_n,t_n)-b(t_n)| \ge \epsilon$. Since $\Omega \times [0,T]$ is compact, without loss of generality, we can assume that $\omega_n \to \omega^* \in \Omega, t_n \to t^* \in [0,T]$. Since $a$ and $b$ are continuous, we deduce that $|a(\omega^*,t^*)-b(t^*)| \ge \epsilon$ and $h(\omega^*) = 1$.

However, $h(\omega^*) = 1$ implies that $\int_0^T{\left|a(\omega^*,t)-b(t)\right|^r}d\mu(t)=0$. Since $\mu$ is absolutely continuous and $a$ and $b$ are continuous, this implies that $a(\omega^*,t)=b(t)$ for all $t\in [0,T]$, which contradicts $|a(\omega^*,t^*)-b(t^*)| \ge \epsilon$.

Therefore, there exists $\alpha <1$ such that $\forall \omega \in \Omega \setminus U$, $h(\omega) \le \alpha$. Hence, by using hypothesis (i), we have
\[
\int_{\Omega \setminus U}{p_n(\omega)\left|a(w,t)-b(t)\right| \,d\omega} \le  \V(\Omega) \delta^n \sup_{(\omega,t)}{|a(\omega,t)-b(t)|}
\]
and hence
\[
\left|\int_{\Omega}{p_n(\omega)a(w,t)d\omega}~-~b(t) \right| \le \epsilon + C_1 \delta^n,
\]
where $C_1$ is a constant that does not depend on $t$ and $\omega$.  Since $\epsilon$ is arbitrary, we deduce that
\[
\lim_{n \to \infty} E_{p_n} [a(\omega, t)] = 
\lim_{n \to \infty}{\int_{\Omega}{p_n(\omega)a(\omega,t)~d\omega}}=b(t)
\]
uniformly in $t \in [0,T]$.  Note that this argument actually shows the somewhat stronger statement that $E_{p_n}[|a(\omega, t) -b(t)|] \to 0$ uniformly in $t$.  The same argument shows that $E_{p_n}[\left|a(w,t)-b(t)\right|^2] \to 0$ uniformly in $t \in [0,T]$.  Hence taking $\bar{a}(t) = E_{p_n}[a(\omega, t)]$, we have 
%

\[
\var_{p_n}\left[a(\omega,t)\right] = E_{p_n}[|a(\omega, t)-b(t)|^2] - |b - \bar{a}(t)|^2
\]
which converges to 0 uniformly in $t \in [0,T]$.
\end{proof}

\section{Numerical examples} \label{sec:examples}
In this section we provide numerical examples to illustrate our theoretical findings and demonstrate the efficacy of our variations on the MINE method to design experiments for dynamics identification.

\subsection{A simple ODE model}
We consider a simple biochemical system that contains 3 chemicals:
\[
A\stackrel{k_1}{\rightarrow}B \stackrel{k_2}{\rightarrow}C
\]
where $k_1$ and $k_2$ are the (unknown) degradation rates of $A$ and $B$, respectively. We also assume that at the beginning, the system only contains $A$.

We model this system using
\begin{align*}
\frac{dA}{dt} =  -k_1A, \ \ \ \frac{dB}{dt} =  k_1A-k_2B, \ \ \  \frac{dC}{dt}  =  k_2B,  
\end{align*}
\[
(A(0),B(0),C(0))=(1,0,0).
\]

In this particular example, we are interested in the dynamics of $B$. The parameter space is $[0.1,10] \times [0.1,10]$, the time interval is [0,180] (seconds) and the ``true'' dynamics of the system will correspond to a fixed value $\omega_0$ that is chosen randomly from the uniform distribution on the parameter space.

The experiments are designed sequentially using criteria \eqref{equa1} or \eqref{equa2}, depending on the assumptions for a given example. In all cases, at step $n+1$, the expected dynamics and the corresponding variance function are calculated using the Markov Chain Monte Carlo method.  A Markov chain of length 10000 with respect to the invariant measure $p_n$ is sampled on the parameter space using Griddy-Gibbs sampling \cite{griddy-gibbs}. To speed up the sampling process, a sparse grid interpolant \cite{sparse-grid} is used to approximate the model output. At each point of the chain, the corresponding dynamics is evaluated using the polynomial interpolant. The average of these sampled dynamics is computed to approximate the EDE, and the variance is approximated in a similar manner.  The interpolant we used in this example has an estimated $L^{\infty}$ error of order $10^{-4}$, which is small in comparison to the experiment error and therefore is negligible. The error of the interpolant is estimated by the difference between the interpolated dynamics and the exact dynamics evaluated using the MatLab solver ode15s at 1000 parameter vectors chosen at random from the uniform distribution on the parameter space.

\begin{figure}
\centering
\begin{subfigure}{.5\textwidth}
  \centering
  \includegraphics[width=1\linewidth]{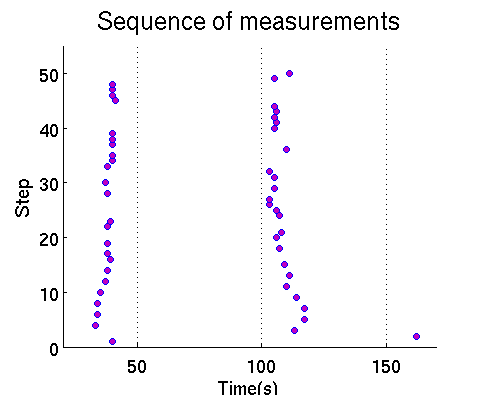}
  \label{fig1a}
\end{subfigure}%
\begin{subfigure}{.5\textwidth}
  \centering
  \includegraphics[width=1\linewidth]{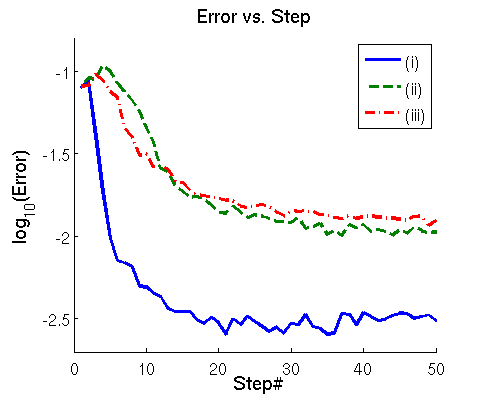}
  \label{fig1b}
\end{subfigure}
\caption{(Two dimensional paramter space) Left: Measured time points designed by MINE criteria. The algorithm focuses on two regions in time that capture the system dynamics. Right: The $L^{\infty}$ errors of EDE on log-scale in three different cases: (i) Data collected with no noise, using the original MINE criteria \eqref{equa1} (ii) Data with Gaussian noise, using the original MINE criteria \eqref{equa1}, (iii) Data with noise, using criterion \eqref{equa2} on a finite set of output values and possible measurement time points. }
\label{Fig1}
\end{figure}

First, we use the framework of Theorem~\ref{Theo2}, in which the data is collected with no noise, the time interval is not discretized, and the experiments are designed using condition \eqref{equa1}. The selected sampling times are shown in Figure~\ref{Fig1}(left panel).  We see that even without the discretization of the possible sampled time points, the algorithm focuses on two regions in time that are sufficient to capture the system dynamics. This is consistent with the fact that the system is controlled by two parameters.  Figure~\ref{Fig1} (right panel (i), solid curve) shows how rapidly the EDE approximates the actual response.  After 5 experiments, the EDE has converged to the true system dynamics within a negligible error.

Next, we consider the case when the data are subject to Gaussian noise with $\sigma^2=0.01$. The dashed curve in Figure~\ref{Fig1} (right panel, (ii)) represents the error of the EDE as described in the original algorithm.  The dash-dot curve (iii) corresponds to the assumptions of Theorem~\ref{Theo5}. In this case, the experiments are designed sequentially using criteria \eqref{equa2} with $C=2$, and the time interval is discretized by a uniform grid whose distance between neighbour points is equal to 20. In either case, the algorithm provides a good approximation of the true dynamics after just a few sequential experiments.

\subsection{An ODE model of the T-cell signaling pathway}

In this example, we consider a mathematical model of the T-cell signaling pathway proposed by Lipniacki et al. in \cite{lipniacki}. This is a system of ODEs with 37 state variables, 19 parameters, and fixed initial conditions.  We seek to design experiments to identify the dynamics of pZAP, one of the state variables of the system.

In this example, the parameter space is defined relative to a nominal parameter vector.  That is, for each component of the nominal vector, we define a range of five times smaller to five times larger than this component.  The whole parameter space is the 19-dimensional set formed by the product of these 19 intervals.  The time interval is [0,201] (seconds). The true dynamics of the system are given by a fixed choice of $\omega_0$ that is chosen randomly from the uniform distribution on the parameter space. The expected dynamics and the corresponding variance function are calculated as described in the previous example. To reduce the computational cost, we also construct a sparse grid interpolant to approximate the output of the ODE system. We use a sparse grid with $50,000$ points to construct the interpolant.  Even so the interpolant has an $L^{\infty}$ error of order $10^{-2}$, so that there is some mismatch in the model.

\begin{figure}
\includegraphics[width=0.6\textwidth]{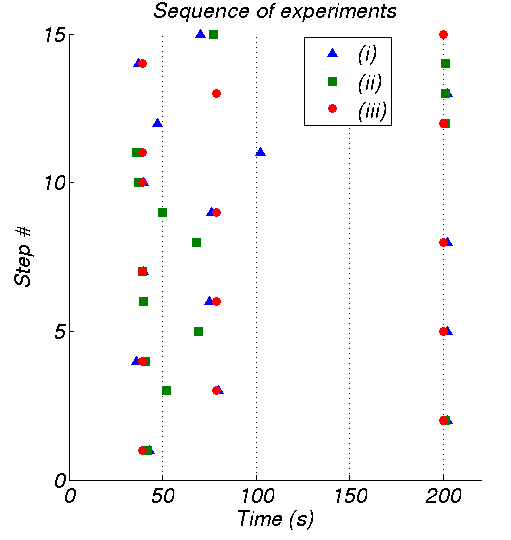}
\caption{(19-dimensional paramter space) Design points in three different cases: (i) Data collected with no noise, using the original MINE criteria (\ref{equa1}) (ii) Data with Gaussian noise, using the original MINE criteria \ref{equa1}, (iii) Data with noise, using criterion \ref{equa2} on a finite set of output values and possible measurement
time points.}
\label{Fig2}
\end{figure}

Figure~\ref{Fig2} shows the sequence of design points created by the algorithm in 3 different cases: (i) Data collected with no noise, using the original MINE criteria \eqref{equa1}; (ii) Data with Gaussian noise, using the original MINE criteria \eqref{equa1}; (iii) Data with noise, using criterion \eqref{equa2} on a finite set of output values and possible measurement time points. In all three cases, the design algorithm focuses on two distinct regions, one of which is precisely defined, the other of which is somewhat nebulous and may perhaps be considered as two regions, particularly in case (iii).  This result suggests that although the ODE system is controlled by 19 different parameters, the set of possible system dynamics is  contained (at least approximately) in a space of dimension 3. { It is worth noting that in \cite{D}, the authors also predicted a pile-up of data points under MINE criteria in the open-loop setting (where multiple measurements are chosen in one step). Our result confirms the same property in the closed-loop case, where the measurements are chosen sequentially with updated probability distributions. }

\begin{figure}
\centering
\begin{subfigure}{.5\textwidth}
  \centering
  \includegraphics[width=1\linewidth]{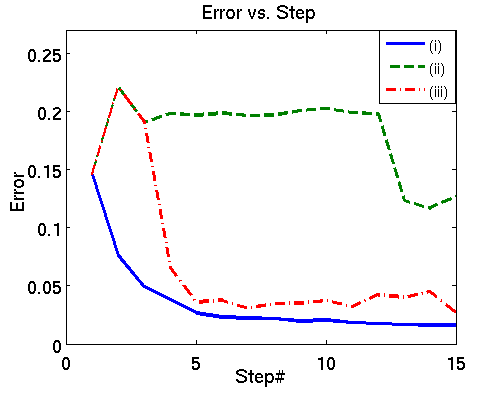}
  \label{fig3a}
\end{subfigure}%
\begin{subfigure}{.5\textwidth}
  \centering
  \includegraphics[width=1\linewidth]{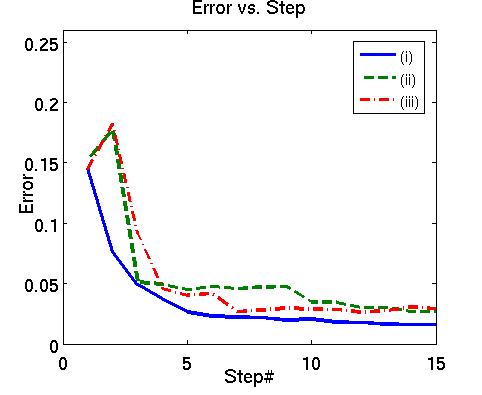}
  \label{fig3b}
\end{subfigure}
\caption{(19-dimensional paramter space). The $L^{\infty}$ errors of EDE on log-scale in three different cases: (i) Data collected with no noise, using the original MINE criteria \eqref{equa1} (ii) Data with Gaussian noise, using the original MINE criteria \eqref{equa1}, (iii) Data with noise, using criterion \eqref{equa2} on a finite set of output values and possible measurement time points. }
\label{Fig3}
\end{figure}

Figure~\ref{Fig3} shows the approximation error of the EDE in these three cases. As in the previous example, the error in case (i) decreases quickly to a level consistent with the error in the interpolant and the error in MCMC sampling. This supports the assertion that if we know the exact values of the dynamics at three important points, we are able to recover the whole time course of the dynamics. 

{
Since our algorithms in case (ii) and (iii) are data-dependent, we present two different realizations of the performance. 

In the first case (left panel of Figure~\ref{Fig3})}, the original algorithm using the MINE criteria with noisy data does not do very well in recovering the dynamics after the first 15 experiments. The problem here is that a measurement with significant noise, especially in the first few steps, can cause the estimator to shift toward a region of parameter space in which the dynamics do not agree with the true dynamics. Moreover, if the output function at this point of measurement is relatively insensitive to parameter changes in this region, it may take many additional measurements to overcome this initial misestimation.

In our example, we encounter this issue: the second measurement made at $t = 201$ gives a data value of nearly 1 when the true value is approximately 0.75. This measurement shifts the probability distribution toward a broad region in the parameter space where the corresponding dynamics saturate to the maximum value 1. This reduces the system variance at time 201 to a relatively small value in comparison to that of other time points. A direct consequence is that in the next eight experiments, no measurement is made around time 200, and the EDE's error does not improve. However, during this process, the parameters that correspond to the true dynamics gain weight, causing the variance around time 200 to increase. Finally, a measurement at time 201 is made in step 11, which significantly decreases the error of the expected dynamics estimator.

This example illustrates the fact that although the convergence of the original algorithm is guaranteed, the convergence may be slow.  Some drawbacks of the original algorithm are removed in case (iii) by replacing criteria \eqref{equa1} by criteria \eqref{equa2} and by restricting the set of possible time points to be finite. By making the set of possible measurement points finite, we collapse the important regions in the time interval to single points and facilitate resampling to get more accurate data at these important points. Also, by using criteria \eqref{equa2}, we obtain the freedom to select the next measurement point subject to multiple criteria, as described next.  

For Figure~\ref{Fig3}, as in the previous example, the set of all possible measurement time points is restricted to a uniform grid of resolution 20, starting from 1. To design experiments, we used the following ranking: among time points with relatively high variance (specifically, that have variance larger than half of the maximum), the time points that have already been measured are given more priority (to promote resampling); among time points that have been measured, the points with fewer measurements have more priority; among time points that have the same number of measurements, the ones with higher variance have more priority.  

The advantages of this variation of the algorithm are illustrated in Figure~\ref{Fig3} (left panel). After 2 measurements that coincide with those of the previous case (including the point with large measurement error), the algorithm then selects a different measurement point that leads the error to drop quickly.  After 6 experiments, the expected dynamics estimator has converged to the true dynamics within an acceptable error. 

{
On the right panel of Figure 3, we consider a different realization of data on the first measurement. In this case, the random data is obtained with small error and leads to quick convergence of the EDEs corresponding to both criteria.
}

\begin{figure}
\centering
\begin{subfigure}{.5\textwidth}
  \centering
  \includegraphics[width=1\linewidth]{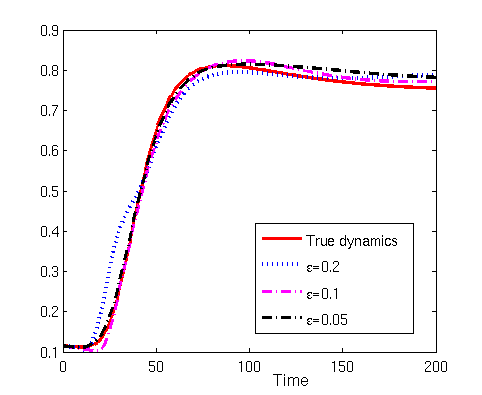}
  \label{fig4a}
\end{subfigure}%
\begin{subfigure}{.5\textwidth}
  \centering
  \includegraphics[width=1\linewidth]{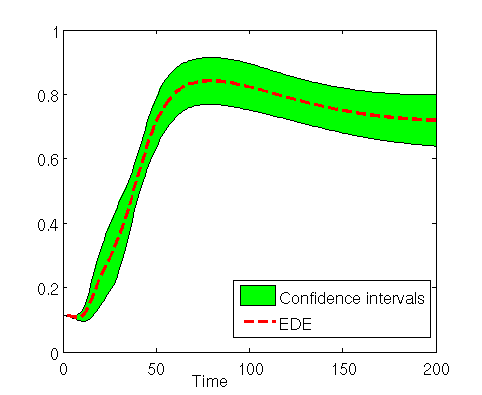}
  \label{fig4b}
\end{subfigure}
\caption{(19-dimensional paramter space). Left: EDEs using different sparse grid interpolators to approximate the dynamics. The EDEs are evaluated after 10 steps. Right: Expected dynamics estimator and predicted confidence intervals of the output dynamics with $\epsilon=0.05$}
\label{Fig4}
\end{figure}

{ This example also illustrates the fact that the  probabilistic framework in experimental design works well in the case when the number of data is less than the number of parameters, or when the model is unidentifiable: our examples couldn't have been done using a method of parameter estimation via optimization. Assume that in example 2, we can make measurements with high accuracy at 3 time points 50, 100 and 200 and want to know the value at time 150. The number of data points in this case is less than the number of parameters and any method that returns a single parameter estimate will never be able to predict with high confidence (or any confidence at all) the output value at 150. In order to do so, it needs to compute every possible parameter values that fit the data, which is very unlikely in practice. Our probabilistic framework provides a feasible way to address the issue: we considered such an example in Figure~\ref{Fig4}, in which we quantify the uncertainty of the dynamics with only 10 noisy measurements ($\sigma=0.1$) that accumulate at 3 time points (see also Figure~\ref{Fig2}), which is much less than the number of model parameters (19).}

{\bf Model mismatch:}
Finally, we illustrate the effect of model mismatch on the convergence of the EDE by using different sparse grid interpolants to approximate the system output. In this particular example, we run the algorithm with the relaxed MINE criteria on a finite set of measured time points and output values with three different sparse grids of 1000, 2000, 9000 grid points, respectively. The estimated $L^\infty$-errors 
of the three interpolants are 0.2, 0.1 and 0.05. As in the previous example, the errors of the interpolants are estimated by the difference between the interpolated dynamics and the exact dynamics evaluted using the MatLab solver ode15s at 1000 parameter vectors chosen uniformly at random from the parameter space. These interpolants are considered as approximate models with varying degrees of mismatch. In each case, the EDE is evaluated after 10 points selected according to criteria \eqref{equa2} with $C=2$. 

The results of this example are given in Figure~\ref{Fig3}. It is not surprising that all three cases give good estimates of the true dynamics: since we are not concerned with the identification of parameters, as long as the dynamics space of the approximate models are close to the dynamics space of the true model, the algorithm will work well. Although the sparse grid interpolant with 1000 grid points is not a good approximation of the system output, it has enough degrees of freedom to capture the behaviour of the system so that a weighted average over parameter space gives a good estimation of the true dynamics.

\section{Conclusion}
Building upon the Maximally Informative Next Experiment algorithm, we have developed several variants of a model-based experiment design algorithm.  This algorithm uses existing data to produce a probability distribution on parameter space and then identifies possible measurement points whose output values have large variance under this distribution.  We have also proven the convergence of the associated EDE (expected dynamics estimator) to the true system dynamics under a variety of assumptions on the model and data, even when the chosen experiments cluster in a small finite set of points.  
{
This approach provides an effective way to incorporate the knowledge arising from nonlinear models into the experiment design process. We illustrated our results with numerical examples on various models of cellular processes.  
}

{
There are several avenues for future work. First, in \cite{D}, the authors proposed several MINE criteria for experimental design. In this work, we establish the theoretical foundations for one of them. The next step would be validating other MINE criteria within a more general model setting.

Second, in this paper, we focused on the identification of observable outputs and did not attempt to address the extrapolation problem, in which measurements of one output are used to make inference about an unobservable output. However, it is worth noting that our framework can be naturally extended to identify unobservable outputs that are theoretically identifiable given that all information about possible observable outputs is known. The problem of determining which unobservable outputs are identifiable in a given experimental setting  will be addressed in one of our independent but related works.
}

\newpage


\newpage

\end{document}